\documentclass[%
 reprint,
superscriptaddress,
 amsmath,
 aps,
prb,nofootinbib
]{revtex4-2}

\usepackage[T1]{fontenc}
\usepackage{lmodern}   
\usepackage{graphicx}
\usepackage{dcolumn}
\usepackage{float}

\usepackage{multirow}
\usepackage{xfrac} 
\usepackage{xcolor}
\usepackage[sort&compress]{natbib}
\usepackage{float}
\usepackage{microtype}

\usepackage{titlesec}

\setcitestyle{square}

\usepackage{hyperref}

\titlespacing\section{0pt}{12pt plus 4pt minus 2pt}{0pt plus 2pt minus 2pt}
\titlespacing\subsection{0pt}{12pt plus 4pt minus 2pt}{0pt plus 2pt minus 2pt}
\titlespacing\subsubsection{0pt}{12pt plus 4pt minus 2pt}{0pt plus 2pt minus 2pt}

\newcommand{\degree}{\ensuremath{^\circ}}

\begin{document}

\title{Intermediate-valence behavior in U$_2$Rh$_2$Sb}
\author{D. Legut}
\email{dominik.legut@vsb.cz}
\affiliation{Department of Condensed Matter Physics, Faculty of Mathematics and Physics, Charles University, Ke Karlovu 3, 121 16 Prague 2, Czech Republic}
\affiliation{IT4Innovations, VSB -- Technical University of Ostrava, 17. listopadu 2172/15, CZ 708 00 Ostrava, Czech Republic}

\author{M. Krnel}
\affiliation{Max-Planck-Institut f\"ur Chemische Physik fester Stoffe, N\"{o}thnitzer Stra{\ss}e 40, Dresden 01187, Germany}

\author{P. Ko\v{z}elj}
\affiliation{Max-Planck-Institut f\"ur Chemische Physik fester Stoffe, N\"{o}thnitzer Stra{\ss}e 40, Dresden 01187, Germany}

\author{Yu.~Prots}
\affiliation{Max-Planck-Institut f\"ur Chemische Physik fester Stoffe, N\"{o}thnitzer Stra{\ss}e 40, Dresden 01187, Germany}

\author{M. Juckel}
\affiliation{Max-Planck-Institut f\"ur Chemische Physik fester Stoffe, N\"{o}thnitzer Stra{\ss}e 40, Dresden 01187, Germany}

\author{U. Burkhardt}
\affiliation{Max-Planck-Institut f\"ur Chemische Physik fester Stoffe, N\"{o}thnitzer Stra{\ss}e 40, Dresden 01187, Germany}

\author{A.~Ormeci}
\affiliation{Max-Planck-Institut f\"ur Chemische Physik fester Stoffe, N\"{o}thnitzer Stra{\ss}e 40, Dresden 01187, Germany}

\author{Yu.~Grin}
\affiliation{Max-Planck-Institut f\"ur Chemische Physik fester Stoffe, N\"{o}thnitzer Stra{\ss}e 40, Dresden 01187, Germany}

\author{A. Leithe-Jasper}
\affiliation{Max-Planck-Institut f\"ur Chemische Physik fester Stoffe, N\"{o}thnitzer Stra{\ss}e 40, Dresden 01187, Germany}

\author{J. Koloren\v{c}}
\affiliation{Institute of Physics (FZU), Czech Academy of Sciences, Na Slovance 2, 182 00 Prague, Czech Republic}

\author{E. Svanidze}
\email{svanidze@cpfs.mpg.de}
\affiliation{Max-Planck-Institut f\"ur Chemische Physik fester Stoffe, N\"{o}thnitzer Stra{\ss}e 40, Dresden 01187, Germany}

\author{U. D. Wdowik}
\affiliation{IT4Innovations, VSB -- Technical University of Ostrava, 17. listopadu 2172/15, CZ 708 00 Ostrava, Czech Republic}

\date{\today}

\begin{abstract} 
Intermediate-valence behavior is sometimes observed in lanthanide-based materials containing Ce, Yb, Sm or Eu. However, the number of actinide-based systems that exhibit this type of behavior is rather limited. In this work, we present the discovery and characterization of a uranium compound U$_2$Rh$_2$Sb, which turns out to be a candidate for the intermediate-valence behavior. The material shows a characteristic feature in the magnetic susceptibility around $T = 50$ K, which can be described within the interconfiguration-fluctuation model of intermediate valence systems. We find the energy difference between the $5f^3$(U$^{3+}$) and $5f^2$(U$^{4+}$) states to be $\Delta E_\text{ex}/k_B \approx 400$ K and the corresponding valence fluctuation temperature to be $T_\text{vf} \approx 140$~K. The value of the electronic specific heat coefficient $\gamma = 50$~mJ mol$_\text{U}^{-1}$K$^{-2}$ signals a modest electron effective mass enhancement. The electrical resistivity indicates metallic behavior, albeit with a small residual resistivity ratio. Measurements of thermoelectric properties indicate a change of sign in the Seebeck coefficient around $T = 100$ K, with a minimum achieved at $T = 50$ K, which coincides
with the broad peak observed in magnetic susceptibility. The experimental results are compared with the theoretical analysis, based on the first-principles calculations, including lattice dynamics.
\end{abstract}

\maketitle
\section{Introduction}

The intermediate-valence behavior in compounds based on lanthanide elements is fairly common \cite{Sampathkumaran_1986, Lawrence_2008,Dilley_1998, Kvashnina_2011, Eggenhoffner_1993, Thompson_1987, Meyer_1992, Tricoire_2021}. However, among actinide-based systems, only a handful of compounds have been suggested to exhibit intermediate valence behavior; see Table \ref{tbl:Uintermvalence} and Refs. \cite{Troc_2006, Troc_2011, Troc_2019, VanDaal_1975, Vitova_2010, Bes_2018, Troc_2007, Robinson_1979, Eggenhoffner_1993,Kohlmann_2000, Noel_1985, Noel_1980, Suski_1980, Thomas_2020,Thompson_1987, Booth_2012}. Similarly to the case of compounds based on lanthanides, the uranium intermediate-valence systems do not order magnetically, having uranium atoms fluctuating between 5$f^3$ (U$^{3+}$) and 5$f^2$ (U$^{4+}$) configurations. By studying these compounds, insights into their complex electronic structure \cite{Hillebrecht_1985, Bes_2018, Kvashnina_2017,Sundermann_2018, Lander_2020, Amorese_2020, Amidani_2021, Yavas_2019, Caciuffo_2010, Havela_2020, Butorin_2021, Ilton_2011} and theoretical description can be obtained.

\begin{table*}
\centering
\caption{\label{tbl:Uintermvalence}A summary of uranium-based
  intermediate-valence systems and of some of their physical characteristics.}
\renewcommand{\arraystretch}{1.5} 
\begin{tabular}{c | c | c | c  c | c  c | c  c |c  } \hline \hline

              &           &           & \multicolumn{2}{c|}{ICF fit \cite{Sales_1975}}      & \multicolumn{2}{c|}{Curie--Weiss fit}    & \multicolumn{2}{c|}{Specific heat parameters} & \multirow{3}*{References}  \\ 
              & $T_{max}$ & $d_{U-U}$ & $\Delta E_{\text{ex}}/k_\text{B}$ & $T_{\text{vf}}$ & $\mu_{\text{eff}}$ & $\Theta_{\text{W}}$ & $\gamma$     & $\Theta_{\text{D}}$            & \\
Units 				& K         & \AA       & K                                 & K               & $\mu_\text{B}$     & K                   & mJ mol$^{-1}_{\text{U}}$ K$^{-2}$ & K         & \\ \hline \hline
U$_2$RuGa$_8$ & 200 & 4.249 & 1080-1430 \textsuperscript{a} & 320-460 \textsuperscript{a} & 3.5-3.8 \textsuperscript{a} & -615-725 \textsuperscript{a} & 57  & 358 &  \cite{Troc_2007, Troc_2006, Schonert_1995, Grin_1988, Troc_2006_2} \\ 
U$_2$Ru$_2$Sn & 175 & 3.557 & 750-1470 \textsuperscript{a} & 180-500 \textsuperscript{a} & 3.3-4.1 \textsuperscript{a} & -520-1440 \textsuperscript{a} & 103 & 183 &  \cite{Mucha_2008, Plessis_2001, Tran_2003, Rajarajan_2007, Tran_2006, Troc_2007, Troc_2006, Mirambet_1994, Troc_2006_2, Strydom_2003} \\ 
URu$_2$Al$_{10}$ & 50       & 5.240 & 293 & 90                                        & 3.3 & $-169$          & 23                                & 384                 & \cite{Troc_2011} \\
U$_2$C$_3$       & 59       & 3.347 & 249 & 59                                        & 2.4 & $-138$          & 40                                & 256                 & \cite{Mallett_1951, Eloirdi_2013} \\ 
U$_2$Rh$_2$Sb    & 50       & 3.648 & 404 & 141                                       & 3.4 & $-452$          & 50                                & 199               & this work \\ 
                 
\hline \hline

\multicolumn{10}{l}{$^{\text{a}}$ \footnotesize Averaged between two crystallographic directions, see Refs. \cite{Troc_2006, Troc_2007} for details.} \\
\end{tabular}
\end{table*}

Among uranium-based systems, some evidence for the existence of intermediate-valence behavior has so far been provided for UM$_2$Al$_3$ (M = Ni, Pd) \cite{Fujimori_1998}, U$_2$Ru$_2$Sn \cite{Troc_2007, Troc_2006, Plessis_2001}, U$_2$RuGa$_8$ \cite{Troc_2007, Troc_2006}, UCoGa$_5$ \cite{Troc_2007},
URu$_2$Al$_{10}$ \cite{Troc_2011, Samsel_2013}, U$_2$C$_3$ \cite{Eloirdi_2013}, U$_3$S$_5$ \cite{Kohlmann_2000}, and UTe$_2$ \cite{Thomas_2020, Fujimori_2021}. In these materials, the magnetic
susceptibility data typically exhibit a broad maximum, which is attributed to the change of the population between the two electronic states of the uranium atoms. Similarly to lanthanide-based materials
\cite{Sales_1975}, the magnetic susceptibility of actinide-based systems can be described within the interconfiguration-fluctuation (ICF) model \cite{Troc_2007, Troc_2006, Troc_2011, Eloirdi_2013},
yielding values of the excitation energy $\Delta E_{\mathrm{ex}}/k_\mathrm{B}$ and the valence fluctuation temperature $T_{\mathrm{vf}}$.
At higher temperatures, Curie--Weiss behavior is typically recovered, from which the effective moment $\mu_{\mathrm{eff}}$ and the Weiss temperature $\Theta_{\mathrm{W}}$ can be extracted.

\begin{figure}
\centering

\includegraphics[trim = 0in 1.1in 6.1in 0in, clip, width=\columnwidth]{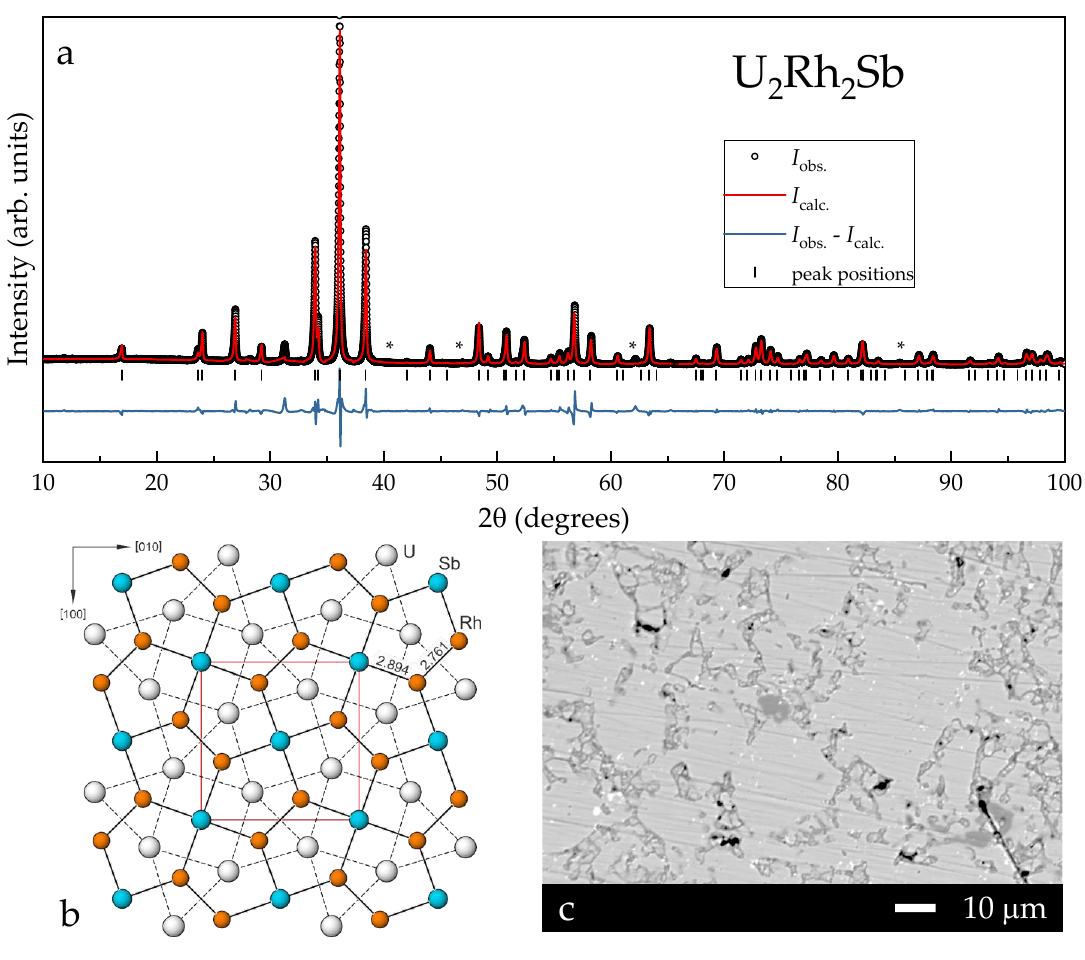}
\caption{(a) Powder x-ray diffraction pattern for U$_2$Rh$_2$Sb with I$_{obs}$, I$_{calc}$, and  I$_{obs}$ - I$_{calc}$, depicted by black circles, red, and blue lines, respectively. The black tick marks represent the calculated peak positions of U$_2$Rh$_2$Sb, while the reflections of the impurity phase are marked with asterisks. (b) Crystal structure of U$_2$Rh$_2$Sb projected along [001]. Alternating planar slabs formed by Rh and Sb atoms at $z = 0$ and U atoms at $z = 1/2$ are highlighted. Dashed lines connect U atoms, yielding CsCl and AlB$_2$ blocks, which  can be used for the description of the crystal structure with the Mo$_2$FeB$_2$ structure type. Shortest interatomic Rh--Rh and Rh--Sb distances are indicated. (c) Microstructure of a U$_2$Rh$_2$Sb sample: U$_2$Rh$_2$Sb (light gray), cavities in U$_2$Rh$_2$Sb (dark gray), holes (black), and impurities (white).}
\label{Xrays}
\end{figure}

In this work, we present the synthesis and characterization of a new compound U$_2$Rh$_2$Sb with structure type Mo$_2$FeB$_2$ -- an ordered variant of U$_3$Si$_2$, see Fig.~\ref{Xrays}(b). The U$_2$Rh$_2$Sb
system is found to be a candidate for intermediate valence behavior with $\Delta E_{\mathrm{ex}}/k_\mathrm{B} \approx 400$~K and $T_{\mathrm{vf}} \approx 140$~K. The value of the Sommerfeld coefficient $\gamma = 50.2$~mJ mol$^{-1}_{\mathrm{U}}$ K$^{-2}$ indicates a modest effective mass enhancement. Electrical resistivity measurements reveal metallic character of this material, which is expected based on the non-zero density of states at the Fermi level, as seen in the band structure calculations. In addition, several experimental properties measured here (such as heat capacity, Debye temperature, magnetic properties) were supported and analyzed by means of density functional theory calculations.

\section{Materials and methods} 

All sample preparation and handling were performed in a specialized laboratory equipped with an argon-filled glove box system (MBraun, $p(\text{H}_2\text{O}/\text{O}_2)<{}0.1$~ppm) \cite{Leithe-Jasper_2006}. Polycrystalline samples of U$_2$Rh$_2$Sb were prepared via arc-melting of the constituent elements (U wire, Goodfellow, $>99.9 \%$; Rh foil, Alfa Aesar, $>99.9 \%$; Sb shot, Alfa Aesar, $>99.9 \%$) in a ratio of 2:2:1. The melting temperature of the U$_2$Rh$_2$Sb compound is $T_{\text{m}} = 1289$ \degree C. Therefore, all samples were annealed for 4 days at $T = 1115$ \degree C, slightly below the melting point. Small inclusions of an unidentified secondary phase(s) in the amount of less than 3\% at. can be estimated from powder X-ray diffraction data (Fig.~\ref{Xrays}(a), asterisks) as well as from metallographic analysis (Fig.~\ref{Xrays}(c), white regions). The material does not exhibit any marked air or moisture sensitivity. 

Powder X-ray diffraction was performed on a Huber G670 Image plate Guinier camera with a Ge-monochromator (Cu~K$\alpha_1$, $\lambda$ = 1.54056 \AA). The resulting data are shown in Fig.~\ref{Xrays}(a). Phase identification was performed using WinXPow software \cite{Stoe_2001}. The lattice parameters (Table~\ref{T1}) were determined by a least-squares refinement using the peak positions, extracted by profile fitting (WinCSD software \cite{Akselrud_2014}). 

The chemical composition of polished samples was studied using energy-dispersive X-ray spectroscopy with a Jeol JSM 6610 scanning electron microscope equipped with an UltraDry EDS detector (ThermoFisher NSS7). Semi-quantitative analysis was performed with 25~keV acceleration voltage and $\approx$ 3 nA beam current. The secondary phases are visible on the scanning electron micrograph; see Fig.~\ref{Xrays}(c). Micrographs also indicate porosity of U$_2$Rh$_2$Sb samples. 

Magnetic properties were studied using a Quantum Design (QD) Magnetic Property Measurement System for the temperature range between 1.8 and 300 K and for applied magnetic fields up to $H=7$ T. The specific heat data were collected on a QD Physical Property Measurement System (PPMS). The low temperature specific heat was measured from 0.4 to 10 K under $H = 0$ and $H = 9$ T.  Additional measurements of specific heat were performed between 2 and 300 K without external magnetic field.  Simultaneous measurements of the Seebeck coefficient $S$, the electrical resistivity $\rho$, and the thermal conductivity $\kappa$, were made using the thermal transport option of the QD PPMS for 2 K $\leq T \leq$ 375 K. Copper leads were attached to the bar-shaped sample using silver conductive adhesive.

Theoretical analysis was based on the density functional theory implemented in the projector-augmented plane-wave (PAW) basis \cite{paw1,paw2} with the PBE-GGA exchange-correlation potential \cite{pbe1}, spin-orbit (SO) coupling, and an additional Coulomb interaction added to the $5f$ shells on the uranium atoms \cite{vasp1,vasp2}, the latter being parametrized by Hubbard repulsion $U$ and  Hund exchange $J$ \cite{Lichtenstein}. The double-counting correction entering the DFT+$U$ functional is taken in the fully localized limit.

Moderate values of $U = 2$ eV and $J = 0.5$ eV were set to adjust the character of the $5f$ states somewhere in the middle between localized and itinerant \cite{Wdowik2016, Wdowik2021,Wdowik2025}. In addition, fully itinerant uranium $5f$ electrons were also considered by setting $U = 0$~ eV and $J = 0$  eV (plain PBE-GGA calculation). Reference configurations of 14, 9, and 5 valence electrons for the U, Rh, and Sb atoms were used. Calculations were performed with an energy cut-off 330 eV for the plane wave expansion. We have considered ferromagnetic (FM) and two types of antiferromagnetic (AF-1 and AF-2) arrangements of the magnetic moments on U atoms \cite{AF-2}, which are schematically shown in Fig.~\ref{fig:magnetic_structures} (Appendix \ref{Appendix-SI}).

The Brillouin zones of the FM and AF-1 structures were sampled with a $8\times8\times16$ $k$-point mesh generated according to the Monkhorst and Pack scheme, while a $8\times8\times8$ mesh was used for
the antiferromagnetic structure AF-2. All structures were fully optimized (lattice parameters and atomic positions) with convergence criteria for total energy and forces acting on each atom to be smaller than $10^{-7}$ eV and $10^{-5}$ eV/\AA, respectively. The lattice contribution to the overall heat capacity has been determined within the harmonic approximation from the calculated density of the phonon states \cite{Parlinski1997,Togo2015,Grimvall}.

\section{Results and discussion}
\subsection{Crystal structure}

The structure of U$_2$Rh$_2$Sb is tetragonal -- space group $P4/mbm$, $a = 7.4011(2)$ \AA, $c = 3.7626(4)$ \AA~ (Rietveld refinement, $R_F = 0.028$, $R_P = 0.060$). With two formula units per cell ($Z = 2$), it resembles a structural motif of Mo$_2$FeB$_2$, being an ordered variant of U$_3$Si$_2$ \cite{Zachariasen_1949}, which are adopted in numerous binary and ternary phases \cite{Lukachuk_2003}. In the aristotype structure, the U atoms are distributed in the two crystallographic positions $2a$ and $4h$, the Si atoms occupy the $4g$ Wyckoff site. One can distinguish two groups of the ternary structures according to the location of the atoms in the Wyckoff positions $2a$ and $4g$. In one group, the position in the tetragonal ($2a$) and trigonal prisms ($4g$) are occupied by the transition metal and the main group element, respectively. This motif is realized in the Mo$_2$FeB$_2$ structure, for instance \cite{Rieger_1964}. The second group has the opposite distribution: the transition metal and the main group element are located at the $4g$ and $2a$ sites, respectively. The latter arrangement is adopted by U$_2$Rh$_2$Sb and was initially discovered for U$_2$Co$_2$Al \cite{Sampaio_1968}. 

\begin{table}
\centering
\caption{\label{T1} Atomic coordinates and isotropic displacement parameters (in \AA$^2$) in the crystal structure of U$_2$Rh$_2$Sb.}
\renewcommand{\arraystretch}{1.5} 
\begin{tabular}{c c c c c c} \hline \hline
Atom & Site & $x/a$        & $y/b$   					  & $z/c$ 			 & $U_{iso}$   \\ \hline
U    & $4h$ & $0.1742(5)$  & $x +{}$\sfrac{1}{2} & \sfrac{1}{2} & $0.025(2)$ \\ 
Rh   & $4g$ & $0.3682(10)$ & $x +{}$\sfrac{1}{2} & $0$ 				 & $0.027(4)$ \\ 
Sb   & $2a$ & $0$          & $0$        				& $0$   			 & $0.034(5)$ \\ \hline \hline
\end{tabular}
\end{table}

In the structure of U$_2$Rh$_2$Sb, planar slabs formed by Rh and Sb atoms alternate with monoatomic U layers along the [001] direction (Fig.~\ref{Xrays} (b)). The Sb atoms have tetragonal coordination by
Rh species, located at a distance of 2.8940(6)~\AA, which is significantly greater than the shortest Sb--Rh distance (2.51-2.75`\AA), observed in binary rhodium antimonides \cite{Karlsruhe_2009}. The Rh--Rh distance of 2.7611(9) \AA\ is comparable to the Rh--Rh distance of 2.689~\AA\ in the ccp structure of the elemental Rh \cite{Donohue_1974}. Interestingly, the U--Rh distances of 2.7669(5) \AA\ and 2.9618(6) \AA, observed in the structure of U$_2$Rh$_2$Sb, are significantly shorter than the U--Sb distance of 3.3189(3)~\AA. Alternatively, the crystal structure of U$_2$Rh$_2$Sb can be described as an arrangement of tetragonal and trigonal prisms, formed by U atoms (marked by dashed lines in Fig.~\ref{Xrays}(b)), filled by Sb and Rh atoms, respectively (CsCl and AlB$_2$-like blocks). The shortest U--U distance in U$_2$Rh$_2$Sb is $d_\text{U--U} = 3.648$ \AA, which is above the Hill limit \cite{Hill_1970}.

\subsection{Magnetic properties}

\begin{figure}
\centering

\includegraphics[width=\columnwidth]{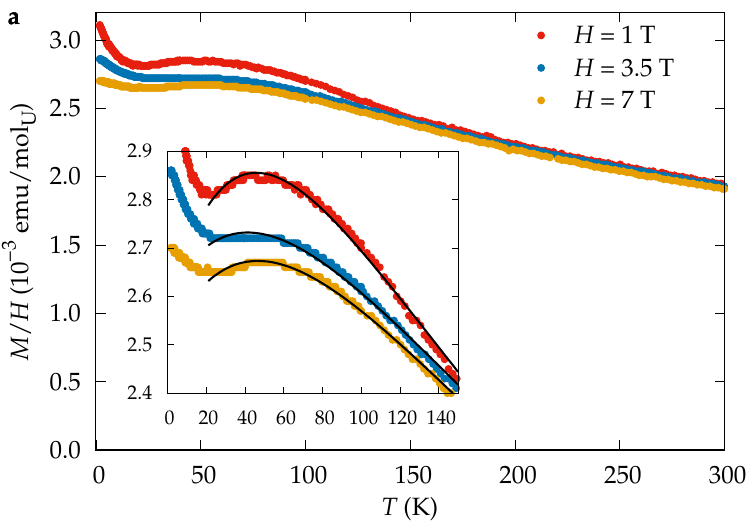}
\includegraphics[width=\columnwidth]{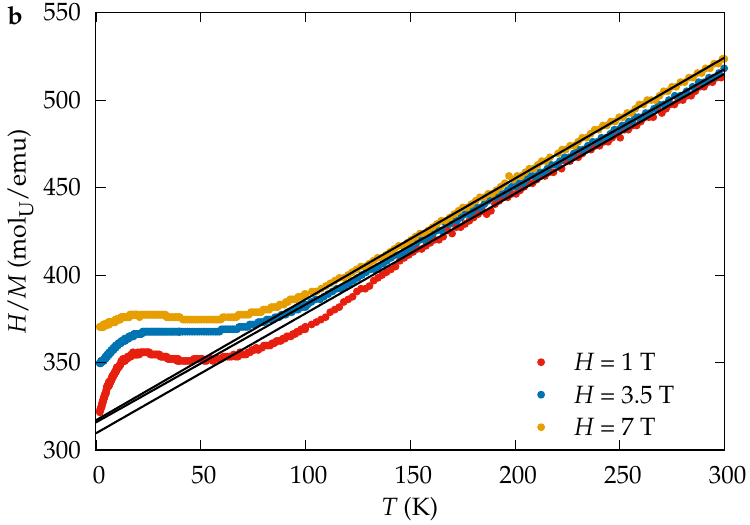}
\caption{(a) Magnetic susceptibility data of U$_2$Rh$_2$Sb in applied magnetic fields $H = 1$~T (red), $H = 3.5$~T (blue) and $H = 7$~T (yellow). Inset: The ICF fits of the data are represented by solid black lines. (b) The inverse magnetic susceptibility data for U$_2$Rh$_2$Sb are fit with the Curie--Weiss law above $T = 150$~K (black lines).}
\label{MT}
\end{figure}

Despite the large separation of the uranium atoms, no ordering has been observed down to $T = 0.35$ K in
U$_2$Rh$_2$Sb. The magnetic susceptibility of U$_2$Rh$_2$Sb, shown in Fig.~\ref{MT}(a), exhibits a broad maximum around $T = 50$ K (see inset). In analogy to materials based on lanthanides, this type of characteristic has been previously attributed to intermediate valence behavior \cite{Troc_2007, Troc_2006, Troc_2011, Eloirdi_2013} using the interconfiguration fluctuation (ICF) model \cite{Sales_1975}, which
assumes two nearly degenerate configurations of the $f$ shell, one with $n$ and the other with $n-1$ electrons. The magnetic susceptibility in this model reads as
\begin{equation}
\chi_{\text{ICF}}(T) = \frac{N_\text{A}}{3 k_\text{B} (T + T_{\text{vf}})} \Big[ \mu_{\text{eff},n}^2 \nu(T) + \mu_{\text{eff},n-1}^2 \big(1 - \nu(T)\big) \Big]\,
\label{eq:susc}
\end{equation}
where $\mu_{\text{eff},n}$ and $\mu_{\text{eff},n-1}$ are the effective magnetic moments of the two configurations, $T_{\text{vf}}$ is the valence fluctuation temperature, and $\nu(T)$ is the canonical occupation of the configuration with $n$ electrons. It can be written as
\begin{equation}
\nu(T) = \frac{2J_{n}+1}{(2J_{n}+1) + (2J_{n-1}+1) e^{ \frac{\Delta
      E_{\text{ex}}}{k_\text{B} (T + T_{\text{vf}})}}}\,,
\label{eq:occ}
\end{equation}
where $J_{n}$ and $J_{n-1}$ are the total angular momenta of the ground-state multiplets of the two configurations, and $\Delta E_{\text{ex}}=E_n-E_{n-1}$ is the energy difference between the two
multiplets. Eqs.~\eqref{eq:susc} and~\eqref{eq:occ} represent a reasonable approximation to the magnetic behavior if $E_{\text{ex}}$ is much smaller than the energy of the excited-state multiplets and, at the same time, if the crystal field splitting of the ground-state multiplets is much smaller than $E_{\text{ex}}$.

The ICF magnetic susceptibility displays a maximum only if the effective moments $\mu_{\mathrm{eff},n}$ and $\mu_{\mathrm{eff},n-1}$ are sufficiently different from each other and if the configuration with the smaller effective moment has a lower energy (and hence it is the state that is preferentially occupied at $T = 0$ K). For the Yb compounds analyzed in \cite{Sales_1975}, this condition is trivially satisfied since Yb fluctuates between $4f^{13}$ and $4f^{14}$ configurations, with 4f$^{14}$ being a singlet with $J_{14} = 0$ and $\mu_{\mathrm{eff},14} = 0$. Uranium, on the other hand, fluctuates between the U$^{4+}$ ($5f^2$) and U$^{3+}$ ($5f^3$) configurations that have nearly identical effective moments, $\mu_{\mathrm{eff,2}} = 3.58 \mu_\mathrm{B}$ and $\mu_{\mathrm{eff},3} = 3.62 \mu_\text{B}$ if the $LS$-coupling scheme is assumed. No susceptibility maximum develops in the ICF model under such conditions. To recover the maximum, some mechanism that would reduce the effective moment of one of the configurations is thus needed.

One scenario, suggested in \cite{Troc_2011}, is crystal-field splitting, which is certainly larger in uranium than in rare earth elements, for which the ICF model was originally devised. Another
effect neglected in the ICF model is the hybridization of the $f$ shell with the rest of the electronic states, which can possibly induce a reduction of $\mu_{\text{eff},2}$ \cite{Eloirdi_2013}. Here, we discuss only possible implications of the crystal-field splitting. Although it is rather straightforward to generalize Eqs.~\eqref{eq:susc} and \eqref{eq:occ} to include the crystal field, the result would not be very useful for the interpretation of our data since it would introduce too many (crystal field) parameters that could not be reliably obtained by fitting the measured $\chi(T)$. For the time being, until we are able to obtain good estimates of the crystal-field parameters by some other means, we follow a simpler, more phenomenological path: We assume that the $J_2 = 4$ multiplet of the 5f$^2$ configuration is split in such a way that there is a singlet ground state separated by a sizable gap from the rest of the multiplet, that is, we have $J^*_2=0$ and $\mu^*_{\text{eff},2}=0$. In that case, the ICF susceptibility simplifies to
\begin{equation}
\chi^*_{\text{ICF}}(T) 
= \frac{N_\text{A} \big(\mu_{\text{eff},3}^*\big)^2}{3 k_\text{B} (T +
  T_{\text{vf}})}
 \frac{2J^*_{3}+1}{(2J^*_{3}+1) + e^{ \frac{\Delta
      E_{\text{ex}}}{k_\text{B} (T + T_{\text{vf}})}}}
 \,,
\label{eq:susc_fit}
\end{equation}
where $J^*_{3}$ and $\mu_{\text{eff},3}^*$ are effective values resulting from the action of the crystal field in the $J_3 = 9/2$ multiplet of the $5f^3$ configuration. These are new fitting parameters in addition to $T_{\text{vf}}$ and $\Delta E_{\text{ex}}$ of the original ICF model.

Fitting all four parameters to the susceptibility data leads to large uncertainties of the fitted values, indicating that the parameters are not independent. To eliminate this problem, we have fixed $J^*_{3}$ to $9/2$, which gives the same fitting function as used in earlier studies \cite{Troc_2007, Troc_2006, Troc_2011, Eloirdi_2013}. An alternative option is to maintain the relation between the effective magnetic moment and the total angular momentum valid in the absence of a crystal field, $\mu_\text{eff,3}^*=g_3 \mu_\text{B} \sqrt{J_3^*(J_3^*+1)}$, where $g_3$ is the corresponding Lande factor. The parameter estimates obtained with this constraint differ from those obtained with $J^*_{3} = 9/2$ only by about 5\% and we do not list them here. 

The high-temperature asymptotics of Eq.~\eqref{eq:susc_fit} read as
\begin{equation}
\chi^*_{\text{ICF}}(T\gg\Delta E_{\text{ex}}) 
\sim \frac{N_\text{A} \big(\mu_{\text{eff},3}^*\big)^2}{3 k_\text{B} (T +
  T_{\text{vf}})}
 \frac{2J^*_{3}+1}{2J^*_{3}+2}
 \,,
\label{eq:susc_fit_asympt}
\end{equation}
that is, it has the Curie--Weiss form with effective moment
\begin{equation}
\mu_\text{eff}=\mu_{\text{eff},3}^*\sqrt{\frac{2J^*_{3}+1}{2J^*_{3}+2}}
\label{eq:mueff_ICF_CW}
\end{equation}
and Weiss temperature $\Theta_\text{W} = -T_\text{vf}$.

Since Eq.~\eqref{eq:susc_fit} applies only at temperatures much lower than the crystal-field splitting of the ground-state multiplets $\Delta_\text{CF}$, the regime described by Eq.~\eqref{eq:susc_fit_asympt} can only be realized when $\Delta_\text{CF}\gg\Delta E_{\text{ex}}$. This condition ensures that there exists a range of temperatures satisfying $\Delta_\text{CF}\gg T\gg\Delta E_{\text{ex}}$. If condition $\Delta_\text{CF}\gg\Delta E_{\text{ex}}$ is not satisfied, the higher crystal-field states start to be thermally populated before the asymptotics of Eq.~\eqref{eq:susc_fit_asympt} can be reached, and the high-temperature Curie--Weiss law ends up having an effective moment different from Eq.~\eqref{eq:mueff_ICF_CW}. The Curie--Weiss regime compatible with Eqs.~\eqref{eq:susc_fit_asympt} and~\eqref{eq:mueff_ICF_CW} was found in URu$_2$Al$_{10}$ \cite{Troc_2011} and U$_2$C$_3$ \cite{Eloirdi_2013}. To satisfactorily fit the high-temperature tail of the magnetic susceptibility of U$_2$RuGa$_8$ and U$_2$Ru$_2$Sn \cite{Troc_2007, Troc_2006} (and also of U$_2$Rh$_2$Sb in the present study), we have to introduce $\mu_\text{eff}$ and $\Theta_\text{W}$ independently of the ICF model. 

The magnetic susceptibility of U$_2$Rh$_2$Sb exhibits an upturn below 20 K, as can be seen from the inset of Fig.~\ref{MT}(a). One possible scenario is that this upturn is caused by a small amount of
magnetic impurity, which can originate from elemental uranium. The ICF fits were performed in the temperature range $20\text{~K} \leq T\leq 150$~K. The upturn below $T = 20$ K cannot be captured within the ICF model, while above $T = 150$ K, the ICF model fails to reproduce the data accurately. The fits in various magnetic fields are shown in the inset of Fig.~\ref{MT}(a) as solid black lines. The values of the excitation energy $\Delta E_{\text{ex}}/k_\text{B}$ and the valence fluctuation temperature $T_{\text{vf}}$ for each of the applied magnetic fields vary slightly, as shown in Table~\ref{tbl:suscfits}. The average values of these parameters are rather similar to those observed in the other candidates for the intermediate-valence behavior based on uranium (Table~\ref{tbl:Uintermvalence}).

\begin{table}
\caption{\label{tbl:suscfits}Magnetic parameters of U$_2$Rh$_2$Sb extracted from the ICF and Curie--Weiss fits at different magnetic fields~$H$. We assumed $J_3^*=9/2$ in the ICF fit.}
\centering
\renewcommand{\arraystretch}{1.5}
\begin{tabular}{c | c | c | c | c |c }
\hline\hline
\multirow{3}*{$H$ (T)} & \multicolumn{3}{c|}{ICF fit}                        		    & \multicolumn{2}{c}{Curie--Weiss fit}           		 \\ 
                       & \multicolumn{3}{c|}{$20$ K $\leq$ T $\leq 150$ K}  			& \multicolumn{2}{c}{$150$ K $\leq$ $T$ $\leq 300$ K}		\\  \cline{2-6}
                       & $\Delta E_{\text{ex}}$ (K) & $T_{\text{vf}}$ (K) & $\mu_\text{eff,3}^*$ ($\mu_\text{B}$)& $\mu_{\text{eff}}$ ($\mu_\text{B}$) & $\Theta_{\text{W}}$ (K) \\ \hline
$1$    & 404 & 141 & 2.82 & $3.42 \pm 0.01$ & $-452 \pm 3$ \\
$3.5$  & 499 & 190 & 3.07 & $3.45 \pm 0.01$ & $-469 \pm 3$ \\
$7$    & 504 & 187 & 3.05 & $3.40 \pm 0.01$ & $-459 \pm 2$ \\
\hline\hline
\end{tabular}
\end{table}

For $T > 150$~K, the inverse magnetic susceptibility data was fit to the Curie--Weiss law. As presented in Fig.~\ref{MT}(b), the fits for the three applied magnetic fields differ only slightly. For $H = 1$ T,
the fit results in an effective moment $\mu_{\text{eff}} = 3.42 \pm 0.01 ~\mu_\text{B}$ and a Weiss temperature $\Theta_{\text{W}} = - 452 \pm 3$ K, which are comparable to those reported for the other
intermediate valence materials based on uranium \cite{Troc_2007,Troc_2006, Troc_2011, Eloirdi_2013}. The value of $\mu_{\text{eff}}$, extracted from the fit, is lower than that expected for U$^{4+}$
($\mu_{\text{eff}}$ = 3.58$\mu_\text{B}$) and U$^{3+}$ ($\mu_{\text{eff}}$ = 3.62$\mu_\text{B}$) configurations, probably due to crystal field splitting in combination with a partial delocalization of the $5f$ orbitals and their hybridization with conduction electrons. Interestingly, the Weiss temperatures $\Theta_{\text{W}}$ for all uranium-based intermediate valence candidates reported are both large in absolute magnitude ($\Theta_{\text{W}} > - 130$ K) and negative, with the maximum absolute value observed in U$_2$Ru$_2$Sn (see Table \ref{tbl:Uintermvalence}). This suggests that perhaps antiferromagnetic correlations are necessary for the observation of the intermediate-valence behavior in uranium-based materials. Furthermore, it appears that ternary uranium-based materials which crystallize in
the Mo$_2$FeB$_2$ structure type are predisposed to antiferromagnetic correlations, given that the majority of U$_2$M$_2$T (M = Fe, Co, Ni, Ru, Rh, Pd and T = Sn, In, Al, Ga) compounds order
antiferromagnetically \cite{Havela_1994, Mirambet_1994, Nakotte_1996,Prokes_2017, Maskova_2019, Mirambet_1993, Strydom_1996, Havela_1995,Bouree_1994, Peron_1993} with the Ne\'el temperature ranging from $T_\text{N} = 15$ K for U$_2$Ni$_2$In \cite{Havela_1994} to $T_\text{N} = 42$ K for U$_2$Pd$_2$Sn. \cite{Tran_1997} This is probably indicative of a strong coupling between crystal structure and magnetism in these materials, as has been shown recently for the U$_2$Rh$_2$Sn system \cite{Gorbunov_2020}. A tendency to antiferromagnetic ordering is also found in our first-principles calculations in Sec.~\ref{sec:abinitio}.

A more in-depth analysis of the intrinsic valence of uranium in U$_2$Rh$_2$Sb would certainly be of great interest. However, it is well known that differentiating between various valence states of uranium is a non-trivial problem, even when advanced spectroscopic tools are involved \cite{Lander_1982,Kvashnina_2017,Caciuffo_2023}. The count of $5f$ electrons in the U atom is never 2 or 3, but rather something in between. In some cases, spectroscopy on a good-quality sample in conjunction with theoretical modeling can be successful in determining the filling of $5f$ states, such as the valence-band HAXPES in combination with DFT + DMFT was in the case of UGa$_2$ \cite{Marino_2024}. A similar analysis of
U$_2$Rh$_2$Sb would be considerably more complex, which is why it was not performed as a part of this work. We hope that further advances in experimental and theoretical tools will address this challenge,
with new materials being crucial for pinpointing specific features responsible for complex valence behavior in actinide-based systems.

\subsection{Heat capacity}
\label{sec:heat_capacity}

\begin{figure}
\centering
\includegraphics[width=\columnwidth,trim={0 0 0 1.3cm},clip]{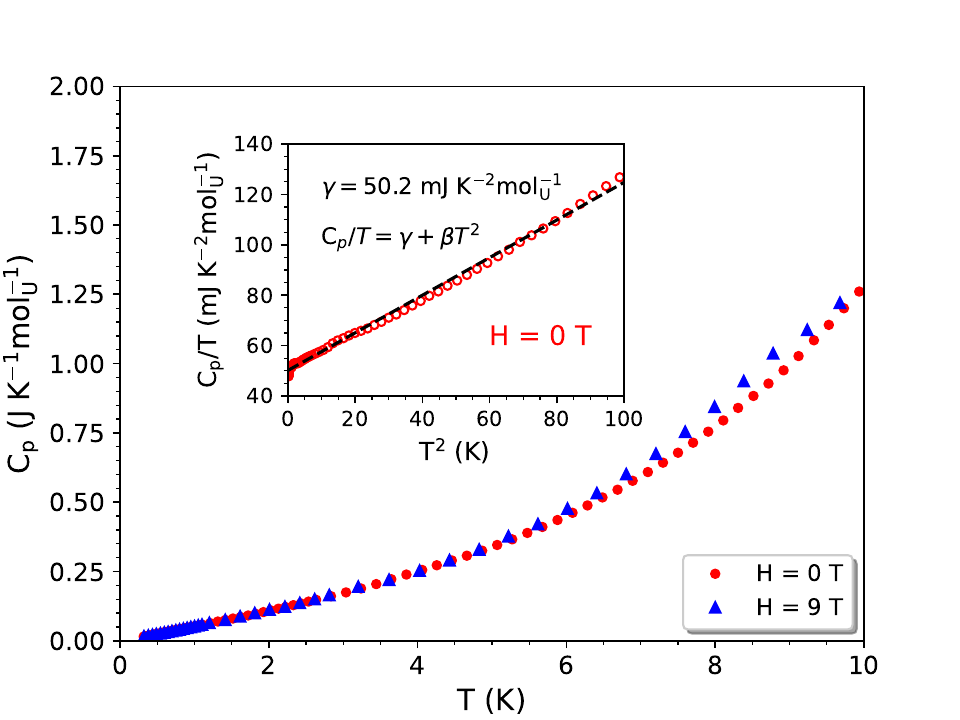}
\caption{Specific heat $C_{\text{p}}$ of U$_2$Rh$_2$Sb measured at $H = 0$ (red symbols) and $H = 9$ T (blue symbols). Inset: Reduced specific heat $C_{\text{p}}/T$ as a function of $T^2$ at $H = 0$. Dashed line stands for the linear fit, from which the value of the Sommerfeld coefficient $\gamma$ is obtained. 
}
\label{Cp_exp}
\end{figure}

The specific heat of U$_2$Rh$_2$Sb measured without as well as with the external magnetic field of 9 T is depicted in Fig.~\ref{Cp_exp}.  The data for $H = 0$ and $H = 9$ T are virtually identical below 10 K. At temperatures above 4 K, the Fermi-liquid behavior of the specific heat can be observed. A linear fit $C_{\text{p}}/T = \gamma+\beta\,T$, indicated by the dashed line in the inset of Fig.~\ref{Cp_exp}, could be used to determine the Sommerfeld coefficient $\gamma = 50.2$ mJ mol$_{\mathrm{U}}^{-1}$ K$^{-2}$. Although the resulting $\gamma$ remains enhanced compared to the value that follows from a free electron model, it is significantly smaller than in U$_2$Rh$_2$In (an isostructural spin-fluctuating system), for which
$\gamma = 140$ mJ mol$_\text{U}^{-1}$ K$^{-2}$ has been reported \cite{Du_1999,Tran_1997, Tran_2006, Nakotte_1994, Havela_1995}. The value of the coefficient $\beta = 0.746$ mJ mol$^{-1}_{\text{U}}$ K$^{-3}$ has been used to estimate the Debye temperature $\Theta_{\text{D}} = 198.5$ K. The estimated $\gamma$ and $\Theta_{\text{D}}$ for U$_2$Rh$_2$Sb are slightly lower than those for U$_2$C$_3$ ($\Theta_{\text{D}} =256$ K \cite{Eloirdi_2013}) and URu$_2$Al$_{10}$ ($\Theta_{\text{D}} = 384$ K \cite{Troc_2011}). To go beyond the approximation of the Debye model in describing the temperature dependence of the heat capacity of URh$_2$Sb$_2$, \textit{ab initio} phonon calculations were undertaken, see Sec.~\ref{sec:abinitio}.

\subsection{Transport properties}

\begin{figure}
\centering
\includegraphics[width=\columnwidth]{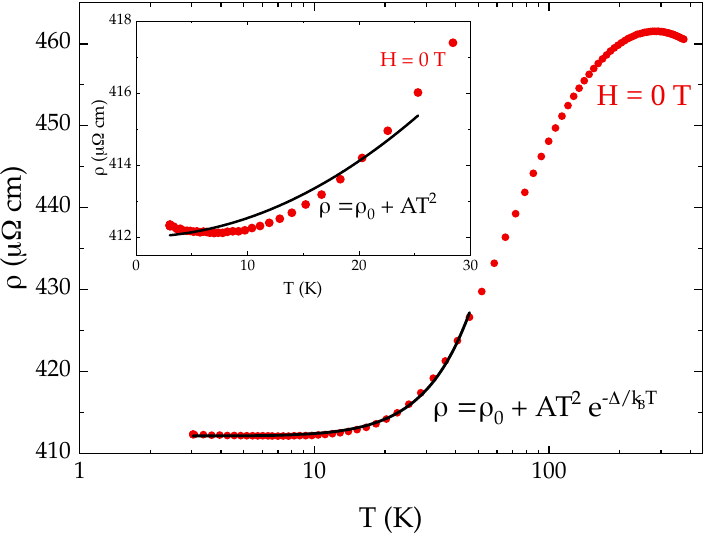}
\caption{Electrical resistivity of U$_2$Rh$_2$Sb in H = 0 T. The inset shows the fit to the Fermi-liquid model, while a fit containing an exponential term is shown in the main panel.}
\label{RT}
\end{figure}

The electrical resistivity data, shown in Fig.~\ref{RT}, indicate metallic behavior, which is consistent with non-zero density of states at the Fermi level as confirmed by the present \textit{ab initio} study discussed in Sec.~\ref{sec:abinitio}. A modest value of the residual resistivity ratio (RRR $= \rho_{300\,\text{K}}/\rho_{2\,\text{K}} = 1.1$) is likely to arise due to the polycrystalline nature of the U$_2$Rh$_2$Sb sample. At low temperatures, the behavior of U$_2$Rh$_2$Sb cannot be fit by the Fermi liquid model, as shown in the inset of Fig. \ref{RT}. Instead, it is better described using the expression \cite{Tran_2017_2, Bukowski_2005}
\begin{equation}
\rho (T) = \rho_0 + A T^2 e^{-\Delta/k_B T}\,,
\end{equation}
where $\rho_0$ is the residual resistivity and the second term corresponds to the scattering of the conduction electrons on spin-wave excitations with an energy gap $\Delta$. The values of $\Delta = 12$ K and $A = 0.00928~\mu \Omega$ cm K$^2$ were extracted from the fit shown in Fig.~\ref{RT}. From this, the Kadowaki--Woods ratio \cite{Kadowaki_1986} $r_{KW} = A/\gamma^2 = 0.34 \times 10^{-5}$ $\mu \Omega$ cm (mol K$^2$ mJ$^{-1}$)$^2$ is derived, which is below the value of $1 \times 10^{-5}$ $\mu \Omega$ cm (mol K$^2$ mJ$^{-1}$)$^2$ expected for heavy-fermion materials.

\begin{figure}
\centering
\includegraphics[width=\linewidth]{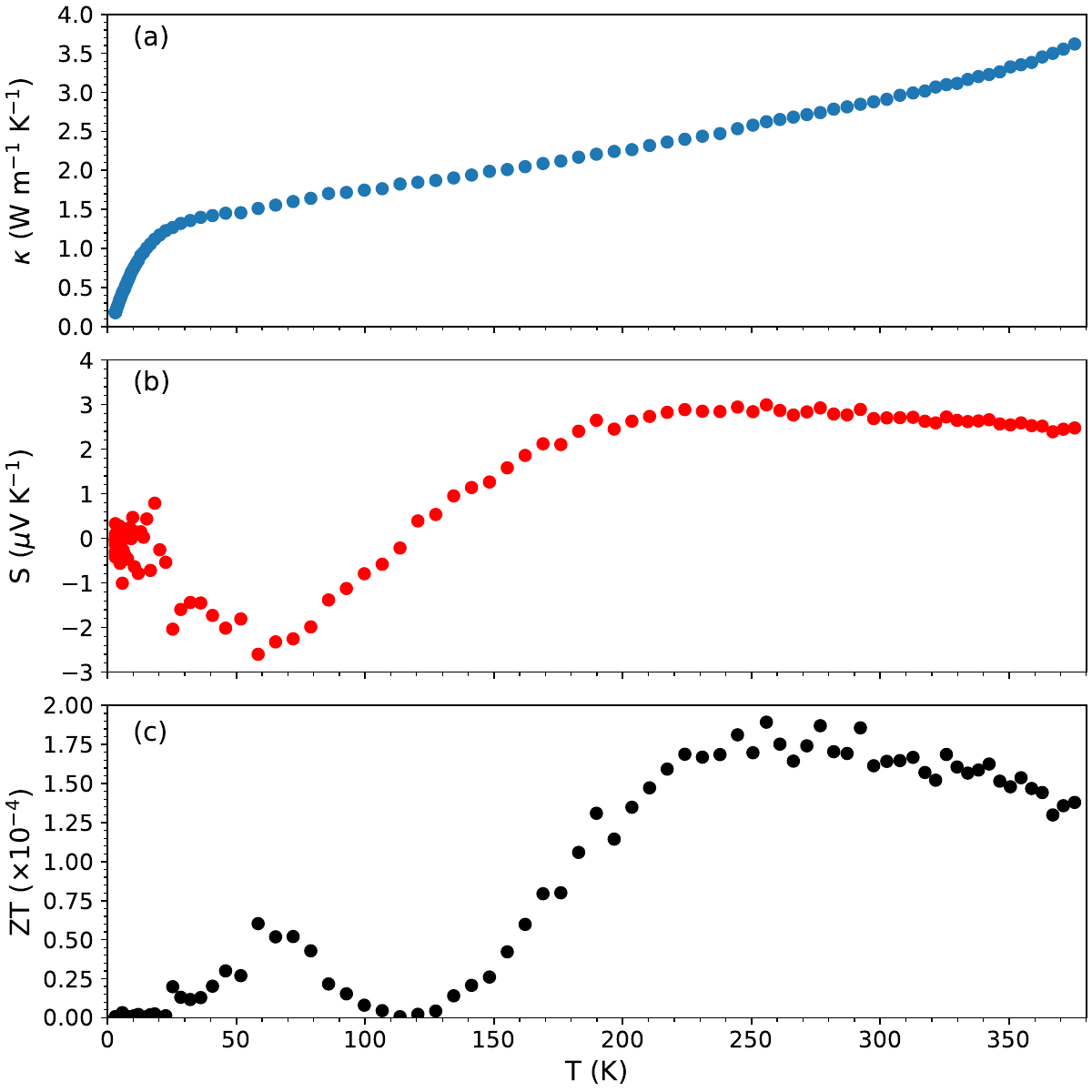}
\caption{Temperature dependence of (a) thermal conductivity~$\kappa$, (b) Seebeck coefficient~$S$, and (c) thermoelectric figure of merit $ZT$ of U$_2$Rh$_2$Sb between 2 and 380 K in zero magnetic field.}
\label{TTO}
\end{figure}

An analysis of the thermoelectric properties of U$_2$Rh$_2$Sb was carried out to compare this material with other compounds based on uranium \cite{Whiting_2018, Svanidze_2019, Freer_2021}. The measured thermal conductivity $\kappa$ and the Seebeck coefficient~$S$, as well as the thermoelectric figure of merit evaluated as $ZT=S^2 \sigma T/\kappa$, where $\sigma$ denotes electrical conductivity, are shown in Fig.~\ref{TTO}. Like in other U-based materials \cite{Troc_2004,Svanidze_2019}, the thermal conductivity in U$_2$Rh$_2$Sb shows a steep increase below 30 K and its temperature dependence up to 50 K can be approximated similarly to typical metallic systems by $\kappa^{-1}= a T^2 + b T^{-1}$, where the first and second terms represent the lattice and impurity scatterings, with $a=15.6 \, \mathrm{m}\, \mathrm{W}^{-1} \, \mathrm{K}^{-1}$ and $b=15.7 \, \mathrm{m}\, \mathrm{W}^{-1} \, \mathrm{K}^{2}$. At temperatures between 50 and 300~K, a linear increase of $\kappa$ is observed with a slope of $5.7\times 10^{-3} \, \mu$VK$^{-2}$. The Seebeck coefficient, which is sensitive to the electronic density of states near the Fermi energy, is negative below $\sim 120$ K and exhibits a minimum of approximately $-2.6 \,\mu$VK$^{-1}$ at 60~K. Above 120~K, the curve $S(T)$ changes sign and the Seebeck coefficient remains positive up to room temperature, where it reaches a broad maximum of $\sim 3 \, \mu$VK$^{-1}$ extending between 200 and 300 K. The overall shape and magnitude of $S(T)$ in U$_2$Rh$_2$Sb are similar to those observed for a number of strongly correlated intermediate valence systems, e.g., URu$_2$Al$_{10}$ \cite{Troc_2011}, U$_2$Rh$_2$In \cite{Tran_2006}, and UCoGa$_5$ \cite{Troc_2004}, and the change in sign of $S(T)$ at low temperatures can be explained to some extent within the theory proposed by Fisher \cite{Fisher_1989}. The temperature variation of the figure of merit is predominantly governed by $S^2(T)$ and thus the $ZT$ curve shows maxima at temperatures corresponding to the minimum and maximum of $S(T)$.  

\subsection{First-principles modeling}
\label{sec:abinitio}

As evidenced by the experimental data shown in the preceding sections, the U$_2$Rh$_2$Sb compound shows paramagnetic (PM) behavior down to very low temperatures. Unfortunately, modeling a PM state is still challenging for first-principles methods, especially for such complex systems as U$_2$Rh$_2$Sb, that require inclusion of spin-orbit interactions and strong on-site correlations to describe not only properties of its electronic and magnetic structures, but also to simulate phonon-dependent properties such as lattice contributions to heat capacity \cite{PM1,PM2,PM3,PM4}. So far, most of theoretical treatment of paramagnetism including contribution from the lattice dynamics effects has been applied to predict behavior of pure metals in their PM states at high temperatures and involved methodologies such as the dynamical mean-field theory (DMFT) \cite{DMFT}, the spin-wave method \cite{SWM} and the spin-space averaging (SSA) formalism within constrained DFT \cite{SSA}. Recently, the disordered local moment model (DLM) and disordered local moment molecular dynamics simulations (DLM-MD) have been quite successful in describing phonons in PM phases of more complex materials \cite{PM5,PM6}. However, such calculations seem hardly feasible for U$_2$Rh$_2$Sb. The most common pathway to depict differences between magnetically ordered and fully disordered systems, which makes use of performing calculations for a hypothetical non-magnetic state with the spin degree of freedom neglected, cannot be considered even as a crude approximation of the PM state, as it may be entirely misleading due to the removal of the magnetic moments from all partially filled atomic shells. Instead, the present \textit{ab initio} calculations have been carried out for ferromagnetic (FM) and two hypothetical antiferromagnetic arrangements (AF-1, AF-2) of spins on U atoms in U$_2$Rh$_2$Sb, as schematically depicted in Fig.~\ref{fig:magnetic_structures} (Appendix~\ref{Appendix-SI}). Such simulations can highlight, to some extent, the importance of spin polarization in predicting the electronic structure and dynamical properties of U$_2$Rh$_2$Sb, regardless of not being fully justified for the description of this system at low temperature.

In the following, we analyze how biasing the U~$5f$ electrons toward itineracy ($U = J=0$ eV) or toward localization ($U = 2$ eV, $J = 0.5$ eV) affects the electronic structure and magnetic properties of U$_2$Rh$_2$Sb in the selected magnetic configurations. Our calculations indicate that among the considered magnetic structures, AF-2 is the most energetically favored compared to the AF-1 and FM solutions, and this holds for both itinerant and localized electrons of the $5f$ shell (cf. Table~\ref{tab:energies} in Appendix~\ref{Appendix-SI}). However, there is some difference in the sequence of energetic stability of the magnetic configurations as the character of the U~$5f$ states varies. We find AF-2 < FM < AF-1 on the itinerant side, which eventually changes to AF-2 < AF-1 < FM when the U~$5f$ electrons become more localized.

The plain GGA functional ($U = J = 0$ eV) substantially underestimates the magnetic moment at the uranium atoms, predicting approximately $M=  0.75\;\mu_B$. The introduction of Hubbard $U$ and the exchange $J$ in the uranium $5f$ states significantly enhances the orbital component of the magnetic moment (see Table~\ref{tab:structure}, Appendix~\ref{Appendix-SI}) and leads to $M=2.4 \; \mu_B$ in the magnetic configuration of AF-2. The DFT-calculated moment $M$ is the maximal projection of the magnetic moment on the quantization axis, which is a different quantity compared to the effective moment $\mu_\text{eff}$ extracted from the Curie--Weiss fit to the susceptibility data. For a lone atomic shell in the $LS$ coupling, the ratio $\mu_\text{eff}/M$ equals $\sqrt{J(J+1)}/J$, that is, 1.1 for two or three $5f$ electrons. The effective moment predicted by GGA+$U$ is hence around $2.65 \; \mu_B$, which is clearly an improvement compared to plain GGA but still noticeably smaller than $\mu_{\mathrm{eff}}$ obtained in the present experiment (3.4 $\mu_\text{B}$).

\begin{figure}
    \centering
    \includegraphics[width=0.99\linewidth,trim={0 0 0 .4cm},clip]{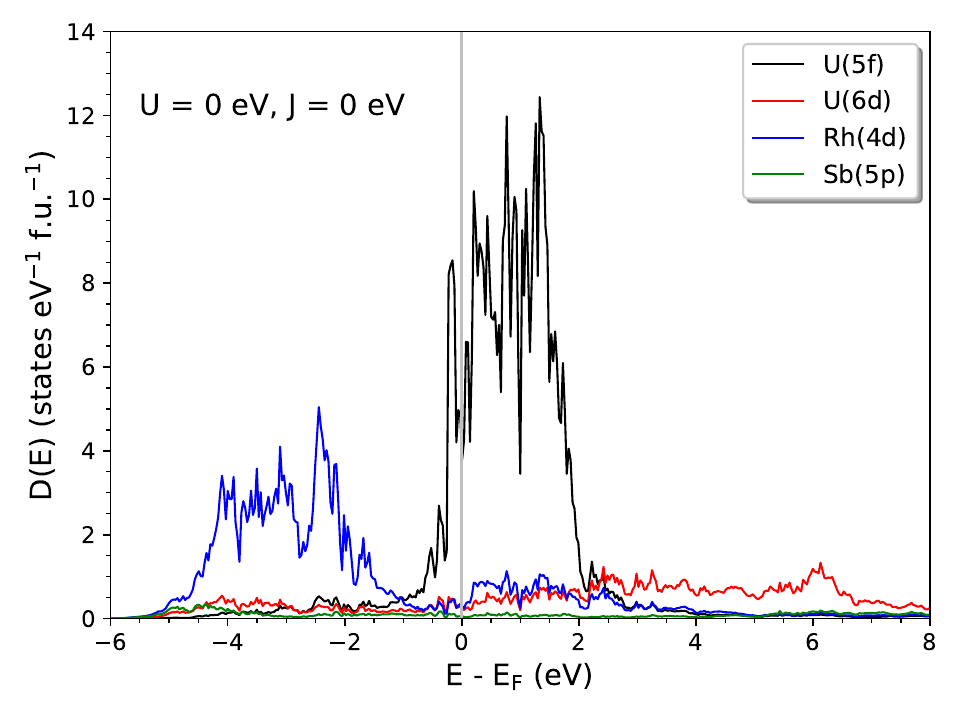}
    \includegraphics[width=0.99\columnwidth]{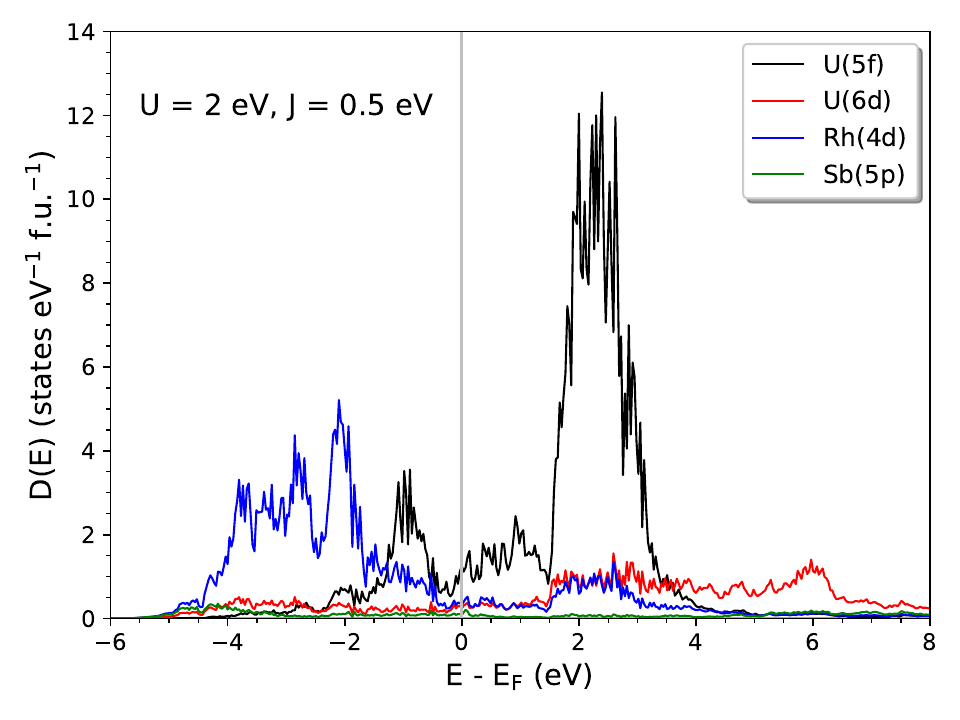}
    \caption{Orbital-projected density of electronic states of the AF-2 antiferromagnetic structure of U$_2$Rh$_2$Sb calculated with spin-orbit interaction and with $U=J=0$ eV (top) or with $U=2$ eV, $J=0.5$ eV (bottom). Projections are shown only for U-, Rh-, and Sb-sublattices corresponding to one spin direction on U-atoms. Energies are given with respect to the reference energy $E_F$ indicated by vertical line.}
    \label{fig:AF2_U2}
\end{figure}

The densities of states of the most stable magnetic phase (AF-2) computed with different values of the Hubbard parameter $U$ ($0$~eV and $2$~eV), are shown in Fig.~\ref{fig:AF2_U2}. They share many common features, among which the most important are (i) a dominant contribution of the U-$5f$ states at the top of the valence band and the bottom of the conduction band, (ii) a large contribution of the Rh-$4d$ states in the energy range between 1 and 5 eV below the Fermi energy and the hybridization of the Rh-$4d$ states with the U-$6d$ states up to about 2.5 eV above the Fermi level, and (iii) a negligibly small contribution of the Sb-$5p$ states at energies ranging from 6 eV below $E_F$ to 6 eV above $E_F$. A variation of the Hubbard parameter induces notable changes only in the U-$5f$ states. At $U=0$~eV, the $5f$ states extend from 1~eV below $E_F$ to 2~eV above $E_F$, whereas they span a wider energy window between 2.5~eV below $E_F$ and 4.5~eV above $E_F$ when additional Coulomb repulsion $U=2$~eV is introduced, causing an increased separation between occupied and unoccupied $5f$ states. This effect is also responsible for a decrease in the density of states at the Fermi energy itself, which leads to a reduced value of the calculated Sommerfeld coefficient whenever the Hubbard parameter is added (see Table \ref{tab:sommerfeld}, Appendix \ref{Appendix-SI}). The computed Sommerfeld coefficient is substantially smaller than the experimental estimate (Sect.~\ref{sec:heat_capacity}), which is expected since the density functional theory neglects many-body renormalizations that enhance the Sommerfeld coefficient of strongly correlated electrons. 

The phonon dynamics of U$_2$Rh$_2$Sb were studied for its AF-2 structure and localized U-$5f$ electrons. The resulting phonon spectrum, shown in Fig.~\ref{fig:phDOS}, covers the frequency range up to about 5.4 THz. It consists of two bands separated by a small gap. The lower band extending to 4.6 THz is composed of three subbands, characterized by the dominating contribution arising from vibrating sublattices of U ($\omega < 2.2$ THz), Rh ($2.2 < \omega < 3.8$ THz) and Sb ($3.8 < \omega < 4.6$ THz). This band mostly involves acoustic and low-lying optical phonon modes. A narrow highest-frequency band ($4.8 < \omega < 5.4$ THz) contains almost pure vibrations of the Rh sublattice and a negligible admixture of vibrations due to U and Sb atoms. 

\begin{figure}
    \centering
    \includegraphics[width=0.99\linewidth,trim={0 0 0 1.7cm},clip]{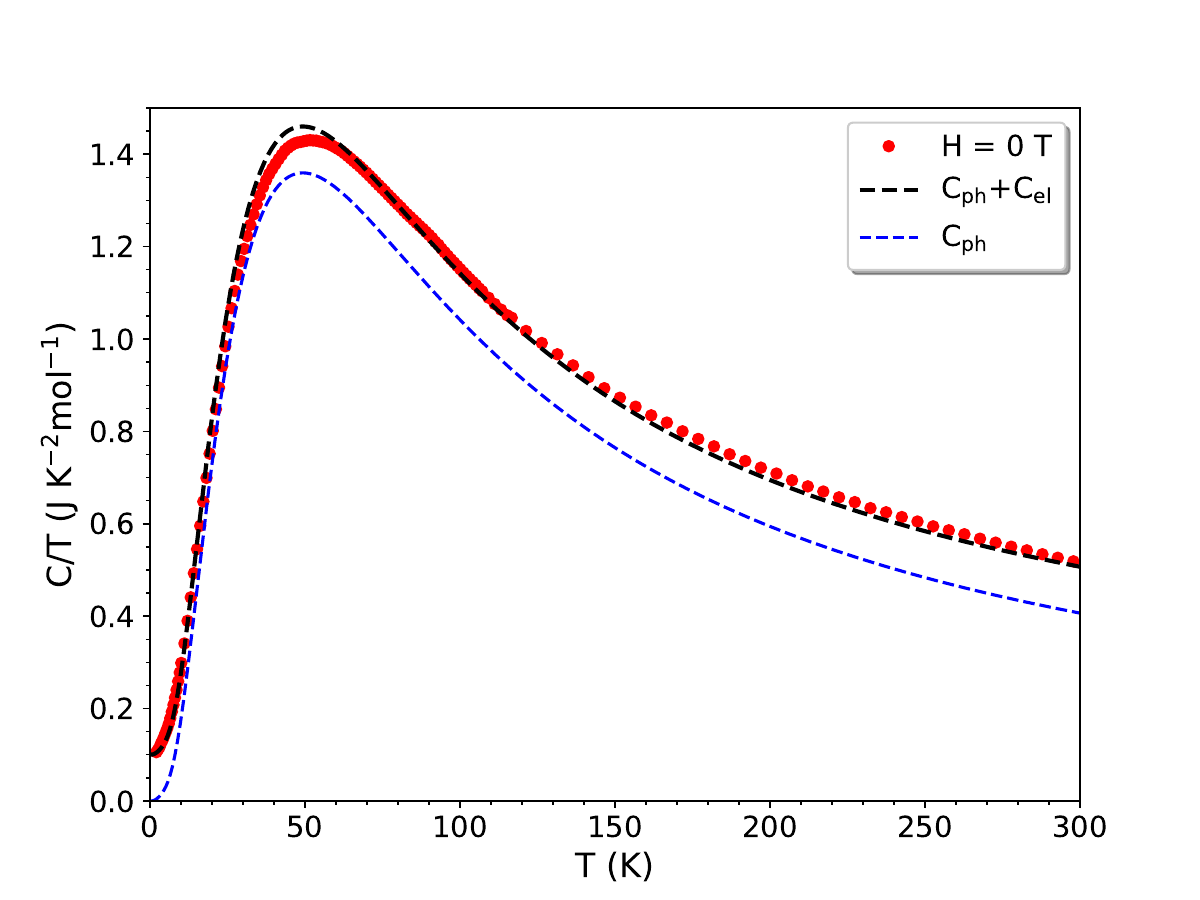}
    \caption{Temperature dependence of the normalized experimental (solid symbols) and theoretical (dashed line) specific heat of U$_2$Rh$_2$Sb. Measurements are carried out from 2 to 300~K without magnetic field ($H=0$~T). Blue dashed line denotes theoretical lattice contribution to heat capacity $C_{ph}$ determined for the AF-2 structure of U$_2$Rh$_2$Sb with $U=2$~eV and $J=0.5$~eV. The total heat capacity $C=C_{ph}+C_{el}$ (black dashed line) is calculated with the electronic term $C_{el}=\gamma T$, where $\gamma = 50.2$ mJ K$^{-2}$mol$_{\mathrm{U}}^{-1}$ is obtained from a fit to the experimental data (Sec.~\ref{sec:heat_capacity}).}
    \label{fig:overall_heat}
\end{figure}

The calculated phonon density of the states has been used to determine the lattice contribution $C_{ph}$ to the overall heat capacity of U$_2$Rh$_2$Sb (see Fig. \ref{fig:overall_heat}). The total heat capacity $C= C_{ph}+C_{el}$, with $C_{el}=\gamma T$ and $\gamma= 50.2$~mJ\,K$^{-2}$mol$_{\mathrm{U}}^{-1}$ determined from the experiment carried out without external magnetic field, follows the experimental specific heat measured between 4 and 300~K very closely. The linear contribution, which describes an additional electronic entropy not captured by the lattice, is significant up to room temperature and remains almost field-independent below 10 K. 

Finally, the second-order moment of the computed phonon spectrum was used to determine the Debye temperature of the AF-2 configuration \cite{Grimvall}. The resulting $\Theta_D = 192.8$ K is somewhat lower, but fairly close to that estimated from the fit of the measured specific heat ($\Theta_D = 198.5$ K, see Sect.~\ref{sec:heat_capacity}). 

\section{Conclusions} 

The interplay between the itinerant and localized nature of the $5f$ electronic states makes their analysis in uranium-based systems inherently difficult. Furthermore, intermediate-valence behavior in uranium-based materials is very rare; so far, it has only been stipulated in a handful of materials. In this work, we present the chemical and physical properties of a newly discovered U$_2$Rh$_2$Sb compound. The intermediate-valence behavior of this system is suggested by the presence of a broad maximum in the magnetic susceptibility around $T = 50$ K. This indicates the existence of nearly energy degenerate 5$f^3$ (U$^{3+}$) and 5$f^2$ (U$^{4+}$) states, which can be described within the interconfiguration-fluctuation model. The values of the excitation energy $\Delta E_{\text{ex}}/k_\text{B} \approx 400$~K and the valence fluctuation temperature $T_{\text{vf}} = 140$~K are similar to those observed in other uranium-based materials. The nonzero density of states at the Fermi level is evident from a moderately enhanced Sommerfeld coefficient $\gamma = 50.2$ mJ mol$^{-1}_{\text{U}}$ K$^{-2}$, metallic character of resistivity, and band structure calculations of U$_2$Rh$_2$Sb. Currently, the features of U$_2$Rh$_2$Sb are consistent with an intermediate-valence behavior, with signs of strong electron correlations at low temperatures. 

Although the presented theoretical model based on density functional theory with inclusion of the spin-orbit interaction and an on-site Coulomb repulsion in the $5f$ shell reasonably approximates the experimentally determined quantities, such as the magnetic moments and the low-temperature heat capacity, the agreement with the experiment is certainly not perfect, and the model remains deficient in several aspects. It is likely that reconsidering U$_2$Rh$_2$Sb as a paramagnet and stepping beyond the static mean-field approximation to the many-body phenomena (which is inherent to the DFT+U method) is necessary to fully understand this compound, but such calculations are still a challenge at the first-principles level. 

Finally, we note that three compounds crystallize with the related U$_3$Si$_2$ structure type: U$_2$Rh$_2$In (spin-fluctuations system \cite{Du_1999, Tran_1997, Tran_2006, Nakotte_1994, Havela_1995}), U$_2$Rh$_2$Sn (antiferromagnet with $T_{\text{N}} = 24$ K \cite{Mirambet_1994, Nakotte_1996}), and U$_2$Rh$_2$Pb (also an antiferromagnet with $T_{\text{N}} = 20$ K \cite{Pospivsil_2020}). A more in-depth joint analysis of magnetic behavior in these and related 221 systems could shed light on the interaction between $5f$ and conduction electrons in these complex materials. Changes in the physical and chemical properties of uranium-based systems crystallizing in the Mo$_2$FeB$_2$ structure have been shown to be easily achieved through chemical substitution \cite{Mirambet_1994, Tran_2000} or hydrogenation \cite{Havela_2006}. This type of investigation would perhaps help to understand the strong interplay between crystallographic properties and magnetism in these materials.

\begin{acknowledgments}
The project No.~25-16339S supported by the Czech Science Foundation and the projects e-INFRA CZ (ID:90254) and QM4ST No. CZ.02.01.01/00/22\_008/0004572 by the Ministry of Education, Youth and Sports of the Czech Republic are acknowledged. ES and MK are grateful for the support of the Christiane Nüsslein-Volhard-Stiftung and the Boehringer-Ingelheim Plus 3 Program. The authors also thank Andrea Severing, Ladislav Havela and Liu Hao Tjeng for fruitful discussions.
\end{acknowledgments}

\bibliography{References}
\newpage
\clearpage

\appendix
\setcounter{figure}{0}
\setcounter{table}{0}

\section{\label{Appendix A} Supplemental Material} 
\label{Appendix-SI}

\setcounter{section}{0}
\makeatletter
\renewcommand{\theequation}{S\arabic{equation}}
\renewcommand{\thefigure}{S\arabic{figure}}
\renewcommand{\thetable}{S\arabic{table}}

\begin{figure}[H]
\centering
\includegraphics[width=0.99\columnwidth]{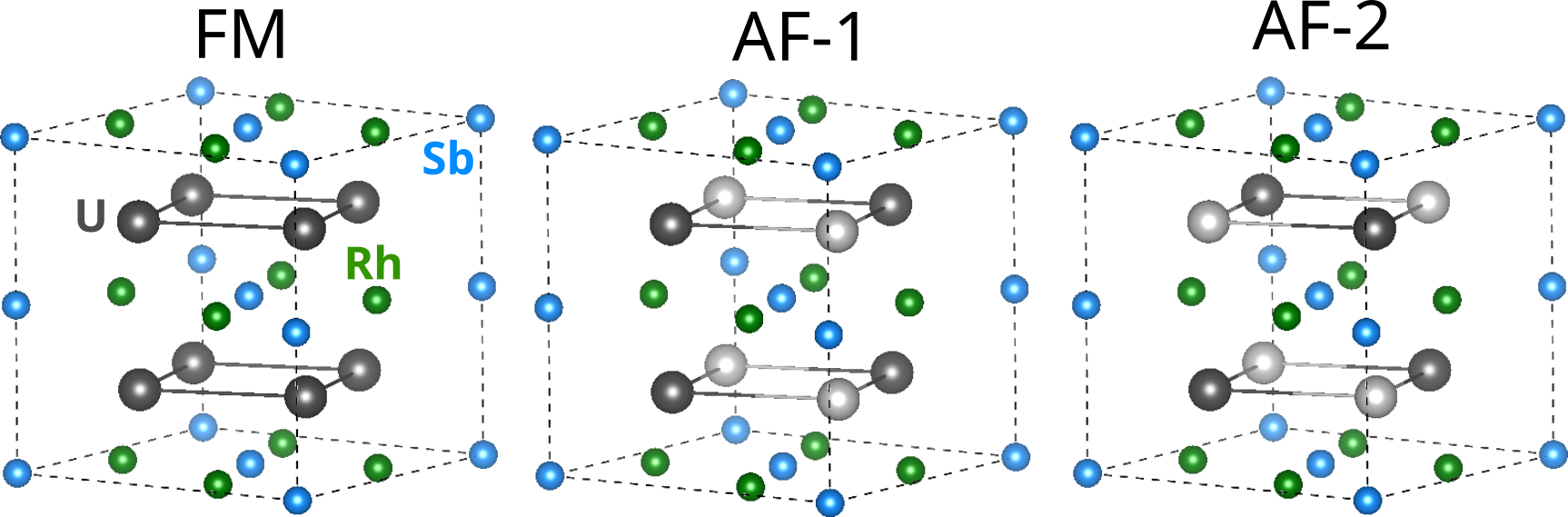}
\caption{Ferromagnetic (FM) and antiferromagnetic (AF-1, AF-2) structures of U$_2$Rh$_2$Sb considered in the present work. Uranium atoms with opposite spin directions are denoted by dark and light gray balls, while green and blue balls indicate Rh and Sb atoms, respectively. The AF-2 magnetic structure has been adopted from Ref.~\cite{AF-2}}
\label{fig:magnetic_structures}
\end{figure}

\begin{table}[H]
\caption{Energies of the AF-1 and AF-2 magnetic structures of U$_2$Rh$_2$Sb with respect to its FM structure taken as reference energy. Units: meV per formula unit.}
\centering
    \begin{tabular}{ccc}
    \hline \hline
      Magnetic  &$U=2$ eV &$U=0$ eV \\ 
      order     &$J=0.5$ eV &$J=0$ eV \\
      \hline
        FM  &   0  & 0    \\
        AF-1 & $-14.9$ & +39.9 \\
        AF-2 & $-26.8$ & $-10.2$ \\
    \hline \hline
   \end{tabular}
    \label{tab:energies}
\end{table}
\begin{table}[H]
\caption{Calculated lattice parameters ($a$, $c$), fractional coordinates of U ($x_{\mathrm{U}}$) and Rh ($x_{\mathrm{Rh}}$) atoms, and spin ($M_S$) orbital ($M_L$) and total ($M$) magnetic moments on U atoms in U$_2$Rh$_2$Sb.}
    \begin{ruledtabular}
    \begin{tabular} {cccccccc}
      \multicolumn{8}{c}{$U=2$ eV\quad $J=0.5$ eV}\\
      \hline\noalign{\vskip 2pt}
      Magnetic   & $a$  & $c$ & $x_{\mathrm{U}}$ & $x_{\mathrm{Rh}}$ & $M_S$ & $-M_L$ & $-M$\\ 
      order      &(\AA) & (\AA) &(--)   &(--)     &($\mu_B$)  &($\mu_B$)  &($\mu_B$) \\
      \hline\noalign{\vskip 2pt}
      FM  & 7.5317 & 3.8290 & 0.1752 & 0.3684 & 1.04 & 3.54 & 2.50 \\
      AF1 & 7.5559 & 3.8034 & 0.1752 & 0.3685 & 1.00 & 3.45 & 2.45 \\
      AF2 & 7.5433 & 3.8174 & 0.1754 & 0.3687 & 1.08 & 3.48 & 2.40 \\
      \hline\noalign{\vskip 2pt}
       \multicolumn{8}{c}{$U=0$ eV\quad $J=0$ eV} \\
       \hline\noalign{\vskip 2pt}
       FM  & 7.4692 & 3.7092 & 0.1738 & 0.3672 & 1.16 & 1.91 & 0.75 \\
       AF1 & 7.4620 & 3.7121 & 0.1736 & 0.3673 & 0.96 & 1.68 & 0.72 \\
       AF2 & 7.4640 & 3.7154 & 0.1743 & 0.3674 & 1.21 & 1.98 & 0.77 \\
       \hline\noalign{\vskip 2pt}
       \multicolumn{8}{c}{Experiment}\\
       \hline\noalign{\vskip 2pt}
       PM & 7.4011 &3.7626 &0.1742 &0.3682 & & \\
    \end{tabular}
    \end{ruledtabular}
    \label{tab:structure}
    \end{table}

\begin{table}[H]
\caption{The Sommerfeld coefficient of FM, AF-1, and AF-2 magnetic structures of U$_2$Rh$_2$Sb, determined in the approximation of spherical Fermi surface as $\gamma=\pi^2k_B^2N(E_F)/3$, where $N(E_F)$ stands for the density of electronic states at the Fermi level. Units: mJ\,K\textsuperscript{$-$2}mol\textsuperscript{$-$1}.}
\centering
    \begin{tabular}{ccc}
    \hline \hline
      Magnetic  &$U=2$ eV &$U=0$ eV \\ 
      order     &$J=0.5$ eV &$J=0$ eV \\
    \hline      
      FM   & 13.2 & 24.0 \\
      AF-1 & 12.7 & 26.8 \\
      AF-2 & 11.4  & 20.3 \\
    \hline \hline
   \end{tabular}
    \label{tab:sommerfeld}
\end{table}

\begin{figure}
\centering
\includegraphics[width=0.99\linewidth]{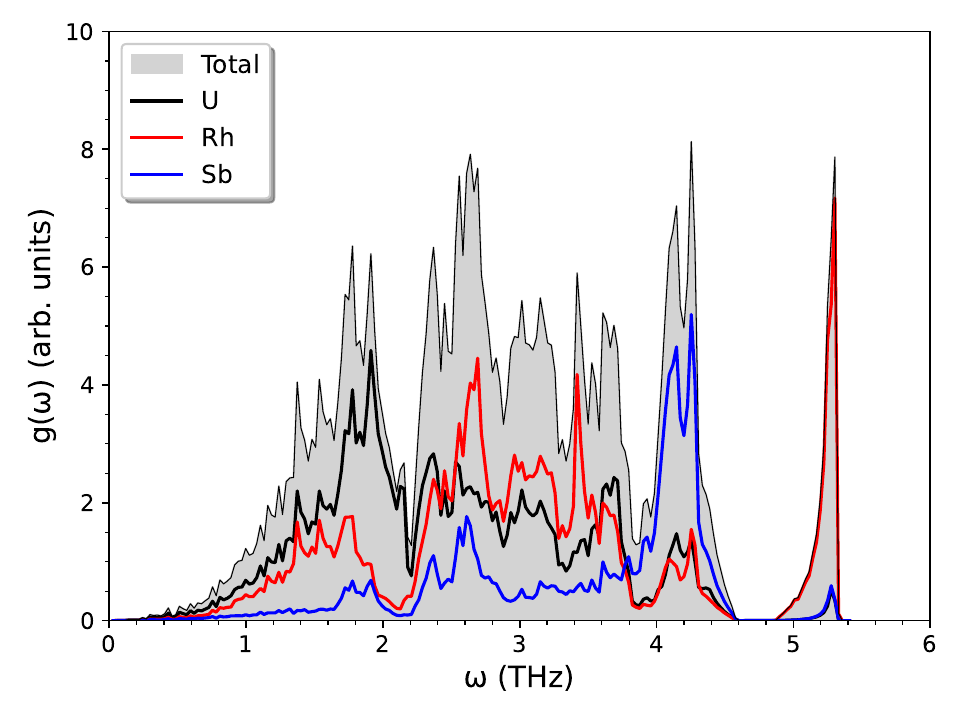}
\caption{Total and partial phonon densities of states for the AF-2 magnetic structure of U$_2$Rh$_2$Sb calculated with $U=2$~eV.}
\label{fig:phDOS}
\end{figure}

\end{document}